\documentclass[a4paper]{article} 

\usepackage[dvipdfmx]{graphicx}
\usepackage[T1]{fontenc}
\usepackage{amssymb}

\newcommand{\argmin}{\mathop{\arg\min}\limits}

\begin{document}

\title{
Instantaneous Spectra Analysis of Pulse Series - Application to Lung Sounds with Abnormalities
}

\author{
Fumihiko Ishiyama
\\
NTT Inc., 
Tokyo, Japan \\
fumihiko.ishiyama@ntt.com
}

\date{}

\maketitle

\begin{abstract}
The origin of the 
``theoretical limit of time-frequency resolution of Fourier analysis''
is from  its numerical implementation, 
especially from an assumption of ``Periodic Boundary Condition (PBC),'' 
which was introduced a century ago.
We 
previously 
proposed to replace this condition with `'Linear eXtrapolation Condition (LXC),''
which does not require periodicity.
This feature makes
instantaneous spectra analysis of pulse series available,
which
replaces
the short time Fourier transform (STFT).
We applied 
the instantaneous spectra analysis
to 
two lung sounds with abnormalities (crackles and wheezing)
and to a normal lung sound,
as a demonstration.
Among them, crackles contains a random pulse series.
The spectrum of each pulse is available,
and the spectrogram of pulse series is available with assembling each spectrum.
As a result, 
the time-frequency structure of given 
pulse series is visualized.
\end{abstract}

%\renewcommand{\thefootnote}{\fnsymbol{footnote}}
%\footnote[0]
%{
%\copyright 2026 IEEE. Personal use of this material is permitted. Permission from IEEE must be obtained for all other uses, in any current or future media, including reprinting/republishing this material for advertising or promotional purposes, creating new collective works, for resale or redistribution to servers or lists, or reuse of any copyrighted component of this work in other works.
%}
%\renewcommand{\thefootnote}{\arabic{footnote}}

\section{Introduction}

\begin{figure}[hbt]
\centering
\includegraphics[width=110mm]{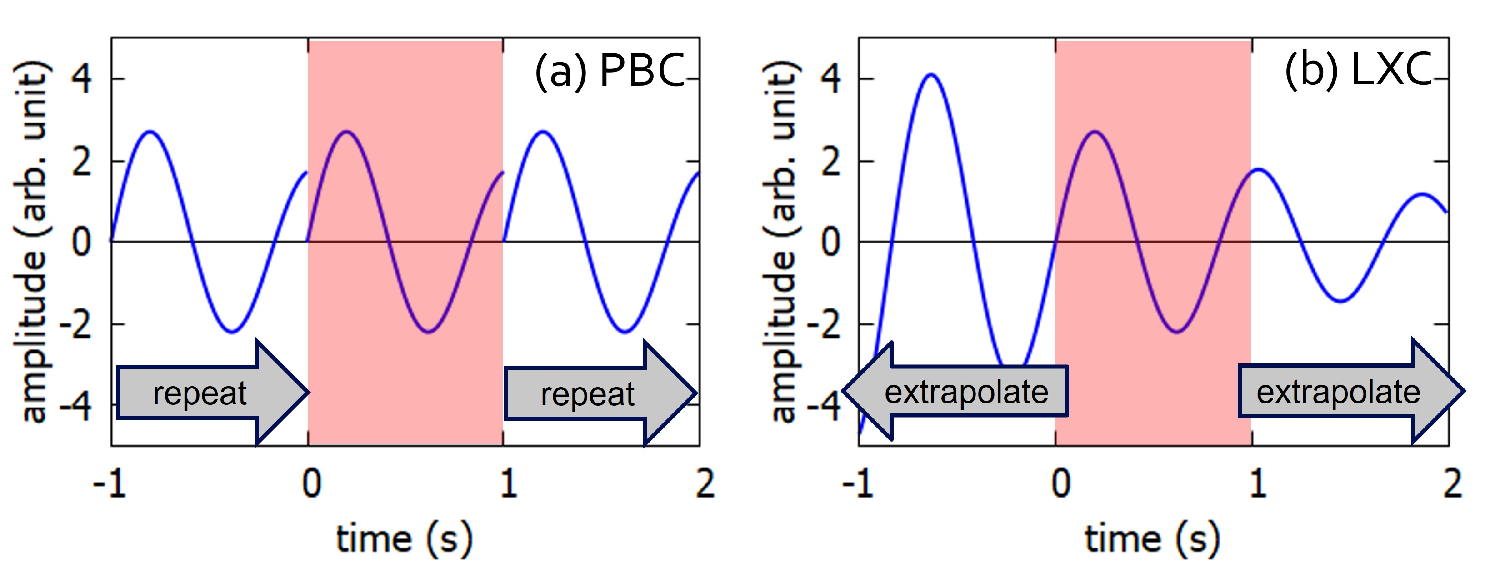}
\caption{
Waveform inside hatched area is given time series for analysis.
(a)~Conventional
Periodic Boundary Condition (PBC), 
which repeats given waveform infinitely \cite{walker}, 
and
(b) 
proposed
Linear eXtrapolation Condition (LXC),
which linearly extrapolates given waveform \cite{cspa2025}.
}
\label{fig-boundaries}
\end{figure}

The origin of the 
``theoretical limit of time-frequency resolution of Fourier analysis''
was thought  to be definite, but it was found that
it is from  its numerical implementation, 
and therefore  it is not  ``theoretical''~\cite{cspa2025,ccisp}.
The point is that the ``Periodic Boundary Condition (PBC),'' 
shown in Fig.~\ref{fig-boundaries}(a),
which is implicitly introduced to the conventional numerical process of Fourier calculation,
and which  implicitly assumes that ``the time series for analysis is infinitely periodic.''
This assumption was required to fulfill the infinite integral for Fourier analysis
\begin{equation}\label{fourier}
S(f)=\int_{-\infty}^\infty s(t) e^{- 2 \pi i f t} {\rm d}t.
\end{equation}

In regard to this matter, 
we previously showed that there is another way~\cite{cspa2025} 
to fulfill the infinite integral,
and the way `'Linear eXtrapolation Condition (LXC)'' 
shown in Fig.~\ref{fig-boundaries}(b) goes beyond the limitation.
It is because the LXC does not require periodicity,
and this feature makes
instantaneous spectra analysis of pulse series available. 

Therefore, we propose to replace
the conventional PBC-Fourier analysis with the LXC-Fourier analysis.
As is shown below, 
the proposed LXC-Fourier analysis theoretically includes 
the conventional PBC-Fourier analysis.

The origin of the PBC will be on the dawn of the quantum mechanics 
around 1930s,
that condition is commonly used such as in the field of solid state physics.
In that historical background,
one of
the first applications of the PBC on numerical analysis
is the work by Walker~(1931)~\cite{walker}, which is the ascending of 
various numerical methods \cite{ccisp} such as the auto correlation, 
linear prediction, 
and
maximum entropy method.

We 
apply the LXC-Fourier analysis to the instantaneous spectra analysis of 
lung sounds with abnormalities
such as ``crackles'' and ``wheezing''
\cite{LS-data},
which are recorded using a digital stethoscope,
as a demonstration.
Among them, ``crackles'' contains a random pulse series. 

As the conventional PBC-Fourier analysis requires infinite periodicity, 
it
is not suitable 
for this 
random pulse series.
In contrast,  proposed LXC-Fourier analysis,
which linearly extrapolates each pulse,
is available.

The spectrum of each pulse is available,
and the spectrogram of pulse series is available with assembling each spectrum.
As a result, 
the time-frequency structure of given 
pulse series is visualized.

In the following sections, 
we introduce the method of LXC-Fourier analysis,
show what are available with the method, 
apply the method to the instantaneous spectra analysis of various lung sounds, 
one of which contains a random pulse series,
and conclude this paper.

\section{LXC-Fourier analysis}

The LXC-Fourier analysis
\cite{cspa2025,ccisp,nolta2023,cqg}
is understood as an expansion of the concept of ``instantaneous frequency''
by van del Pol \cite{pol}.
The concept corresponds to the flame work of frequency modulation (FM),
and we add the concept of amplitude modulation (AM) in it. 
Then, we have a series expansion with AM-FM oscillations,
which is shown below.

Remind that the
conventional PBC-Fourier analysis is a series expansion 
without FM, and also without AM.
Therefore, 
another flame work for 
the LXC-Fourier analysis 
is required.

 In addition, 
 we introduce the concept of ``local linearization'' by Kubo  \cite{kubo},
and we numerically calculate local linearized solution.
This concept is required 
to obtain a unique series expansion~\cite{daubeshies2011,cspa2023,cspa2025b}.

\subsection{Model equation}

We expand the given time series $S(t) \in \mathbb{R}$ with 
general complex functions $H_m(t) \in \mathbb{C}$ as 
\begin{equation}\label{eq-noe}
S(t) = \sum_{m=1}^M e^{H_m(t)},
\end{equation}
where $M$ is the number of complex functions.

The complex functions are expressed as 
\begin{equation}\label{eq-h-fl}
H_m(t) = \ln c_m(t_0) + \int_{t_0}^t [ 2 \pi i f_m(\tau) + \lambda_m(\tau) ] {\rm d}\tau,
\end{equation}
where $f_m(t)~\in~\mathbb{R}$ represents the FM terms,
which is known as instantaneous frequency~\cite{pol} by van der Pol,
$\lambda_m(t)~\in~\mathbb{R}$ represents the AM terms,
which is our original~\cite{nolta2023,cqg},
and $c_m(t_0) \in \mathbb{C}$ represents the amplitudes of the terms at $t=t_0$.

This expansion corresponds to a mode decomposition with general complex functions,
noting that 
\begin{equation}
H_m^\prime(t) = 2 \pi i f_m(t) + \lambda_m(t).
\end{equation}

Additionally, note that
the case
\begin{equation}
H_m^\prime(t) = 2 \pi i \frac{m}{M \Delta T},
\end{equation}
becomes conventional PBC-Fourier series expansion
\begin{equation}
S(t) = \sum_{m=1}^M e^{H_m(t)}=\sum_{m=1}^M c_m(t_0) e^{2 \pi i \frac{m}{M \Delta T} (t-t_0)},
\end{equation}
itself.
That is, our model equation contains PBC-Fourier analysis 
as a special case,
and is a natural expansion.

\subsection{Locally linearized solution}

It is known that this kind of AM-FM series expansion 
does not have a unique solution
\cite{daubeshies2011,cspa2023,cspa2025b}.
Daubeshies presented such an example \cite{daubeshies2011}
with a simple time series
\begin{eqnarray}
S(t) 
&=&
\frac{1}{4} \cos (\Omega-\gamma) t
+
\frac{1}{4} \cos  (\Omega+\gamma) t
+
\frac{5}{2} \cos  \Omega t
\label{eq-threeExp}
\\
&=&
\left(
2+\cos^2  \frac{\gamma}{2} t 
\right)
\cos  \Omega t.
\label{eq-oneAM}
\end{eqnarray}

\begin{figure}[hbt]
\centering
\includegraphics[width=90mm]{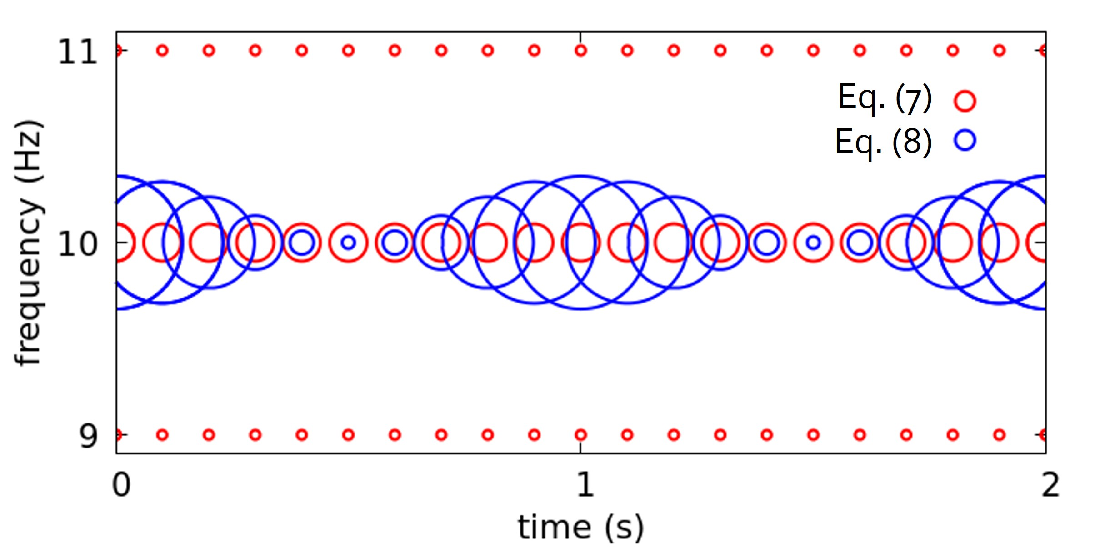}
\caption{
Two spectrograms for single waveform
\cite{cspa2025b}.
}
\label{fig-am-fm-specg}
\end{figure}

It means that there are multiple expressions
for a single waveform,
and each expression has its own spectrogram  \cite{cspa2025b}
shown in Fig.~\ref{fig-am-fm-specg}.

Therefore, we introduce the concept of
``local linearization'' by Kubo \cite{kubo}
to overcome this issue \cite{cspa2023},
which corresponding to selecting Eq.~(\ref{eq-threeExp})
and the LXC-Fourier analysis.

For the purpose, 
we expand Eq.~(\ref{eq-noe}) as 
\begin{equation}\label{eq-exp-h-local}
S(t)|_{t \sim t_k} \simeq \sum_{m=1}^M e^{H_m(t_k)+H_m^\prime(t_k) (t-t_k) + 
O\left((t-t_k \right)^2)},
\end{equation}
around $t \sim t_k=t_0+k \Delta T$,
consider a short enough time width, 
and ignore the higher order terms $O((t-t_k)^2)$.

Then, 
the equation becomes a simple linear equation
\begin{eqnarray}
\label{eq-exp-h-locallinear}
S(t)|_{t \sim t_k} &\simeq& \sum_{m=1}^M e^{H_m(t_k)+H_m^\prime(t_k) (t-t_k) }
\\
\label{eq-exp-h-locallinear-b}
&=& \sum_{m=1}^M c_m(t_k) e^{[2 \pi i f_m(t_k)+\lambda_m(t_k)] (t-t_k) },
\end{eqnarray}
and we can obtain unique $H_m^\prime(t_k)$ easily 
by applying the numerical method of  
linear predictive coding (LPC) with $N$ samples, 
noting that we must use a non-standard numerical method~\cite{ccisp,git}
to hold the condition LXC shown in Fig.~\ref{fig-boundaries}(b).
The standard method of LPC is not adequate,
because it contains
the unfavorable condition PBC by Walker~\cite{walker} 
shown in Fig.~\ref{fig-boundaries}(a),
and
an unfavorable approximation by Itakura~\cite{itakura} to reduce computation cost.

Subsequently, we calculate 
the complex amplitudes $c_m(t_k)$
of the terms
$c_m(t_k) e^{H_m^\prime(t_k) (t-t_k)}$ as
\begin{equation}\label{eq-calc-amps}
\argmin_{c_{m}(t_k)}
\sum_{n=0}^{N-1} 
\left(
S(t_k+n\Delta T)
-
\sum_{m=1}^M c_m(t_k) e^{n H_m^\prime(t_k) \Delta T}
\right)^2.
\end{equation}

%\newpage
\subsection{Instantaneous spectrum}

The equation for instantaneous spectrum is given as follows:
we transform 
Eq.~(\ref{eq-exp-h-locallinear-b})
with a complex integral
as
\begin{eqnarray}\label{eq-specg-cont}
F(f,t_k)&=& \int_C S(t)|_{t \sim t_k} e^{-2 \pi i ft} {\rm d}t 
\nonumber
\\
%%&=& 
%%\int_C \sum_m e^{H_m(t_k)+ H_m^\prime(t_k)t -2 \pi i ft} dt 
%%\nonumber
%%\\
%&=& 
%\sum_m \int_C c_m(t_k) e^{[\lambda_m(t_k)+ 2 \pi i f_m(t_k)]t -2 \pi i ft} dt 
%\nonumber
%\\
%&=& 
%\sum_m c_m(t_k) \int_C  e^{[\lambda_m(t_k)+ 2 \pi i (f_m(t_k)-f)]t} dt 
%\nonumber
%\\
&=& 
\sum_m \frac{ c_m(t_k) }{\lambda_m(t_k)+ 2 \pi i (f_m(t_k)-f)}.
\end{eqnarray}

As Eq.~(\ref{eq-specg-cont}) is for continuous systems, 
some modifications 
to adopt for discrete systems are required.
That is, 
specifically, 
a modification from a Fourier transform to a Fourier series expansion.

For example, when $\lambda_m(t_k) = 0$, 
Eq.~(\ref{eq-specg-cont})  is unbound at $f=f_m(t_k)$, and is not practical.
The practical value is the maximum value $|c_m(t_k)|$ for discrete systems.

Therefore, we take the absolute value of each term, and 
adjust the maximum values at  $f=f_m(t_k)$ so as to be~$|c_m(t_k)|$~\cite{ccisp}.

\begin{equation}\label{eq-specg-disc}
F_{\rm disc}(f,t_k)=\sum_m 
\left| \frac{ c_m(t_k) \lambda_m(t_k) }{\lambda_m(t_k)+ 2 \pi i (|f_m(t_k)|-f)} \right|
\end{equation}

This equation has several merits~\cite{ccisp,nolta2023}. For example,
the instantaneous spectrum of each term is available,
and this feature is valuable for signal separation,
as we show below.
In addition, 
$F_{\rm disc}(f,t_k)^2$ becomes power spectrum, 
corresponding to the conventional Fourier power spectrum.

Another equation for instantaneous spectrum
\begin{equation}\label{eq-specg-disc-pm}
F_{\pm}(f,t_k)=
\sum_m 
\left| \frac{ c_m(t_k)  }{\lambda_m(t_k)+ 2 \pi i (|f_m(t_k)|-f)} \right| \lambda_m(t_k),
\end{equation}
is also available with our method.
Negative intensity becomes available with this equation.
The positive intensity means that the signal is growing ($\lambda_m(t_k)>0$), 
and the negative intensity means that the signal is decaying ($\lambda_m(t_k)<0$).

\subsection{Revised time-frequency resolution}

We briefly demonstrate 
how LXC-Fourier analysis works~\cite{git}.
The source code and its execution output of the followings are shown in the reference.

\begin{figure}[hbt]
\centering
\includegraphics[width=60mm]{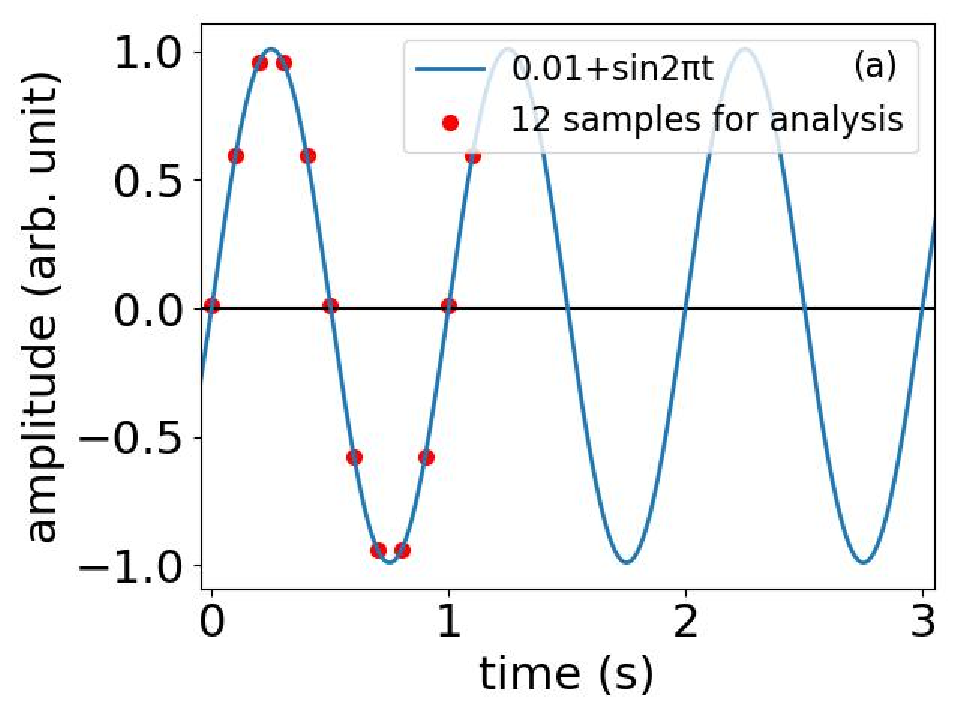}
\includegraphics[width=60mm]{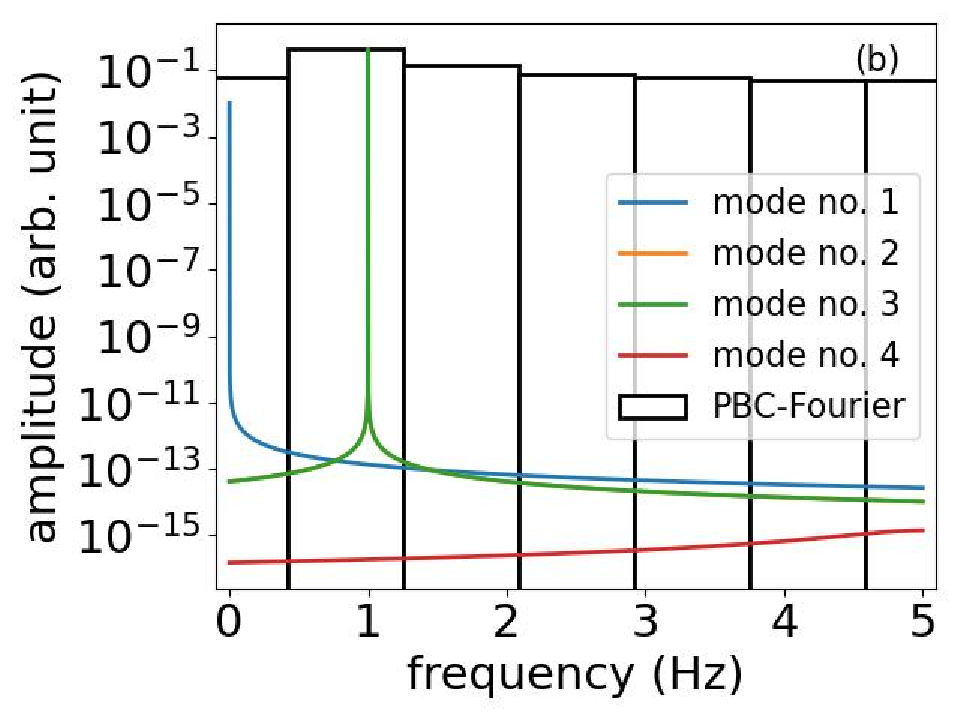}
\caption{
(a) Waveform, and 12 samples for analysis.
(b) Obtained spectra with LXC- and PBC-Fourier analysis \cite{git}.
}
\label{fig-demo}
\end{figure}

The time series for analysis shown in Fig.~\ref{fig-demo}(a) is
\begin{equation}
S(t)= 0.01 + \sin 2 \pi t,
\end{equation}
and we take twelve samples
with a sampling frequency of $10$~Hz,
as shown in the figure.
The samples correspond to $1.2$ cycles of the oscillation.

Following this, we apply the LXC-Fourier analysis to the twelve samples, 
and 
plot each obtained term in Eq.~(\ref{eq-specg-disc})
in Fig.~\ref{fig-demo}(b).

In addition, we plot the PBC-Fourier spectrum from the same twelve samples
in the figure,
which corresponds to a bin of short time Fourier transform (STFT).

Four LXC-Fourier spectra 
(no.~1 to 4)
are shown in Fig.~\ref{fig-demo}(b).
Each spectrum corresponds to 
(no.~1)
a constant term with amplitude $0.01$,
 (no.~2 and 3, the same spectra)
a sinusoidal time series with a frequency of $1$ Hz,
and
 (no.~4)
computational error, 
which corresponds to white noise on the time series.

Note that white noise on the given time series (no.~4) is obtained 
as a mode with a flat spectrum,
and we can remove this unnecessary term
from Eq.~(\ref{eq-specg-disc}) for plotting 
spectrum.

Obtained numerical results for no. 1 to 3 are shown in Table~\ref{tab-tf-resolution}.
\begin{table}[htb]
\caption{numerical resolutions}
\label{tab-tf-resolution}
\centerline
{
\begin{tabular}{c|c|c}
term & frequency (Hz) & amplitude
\\
\hline 
%constant & 
no. 1 &
0.0~~~~~~~~~~~~~~~~~~~~~ &
0.009999999999492205
\\
\hline 
%sinusoidal & 
no. 2 \& 3 &
~0.9999999999999438 & 
~1.0000000000004858~~~~
\\
\hline
\end{tabular}
}
\end{table}
We can obtain this numerical resolution with using twelve samples.
In contrast, the resolution with PBC-Fourier analysis is restricted to $1/1.2$~Hz.

\begin{figure}[hbt]
\centering
\includegraphics[width=110mm]{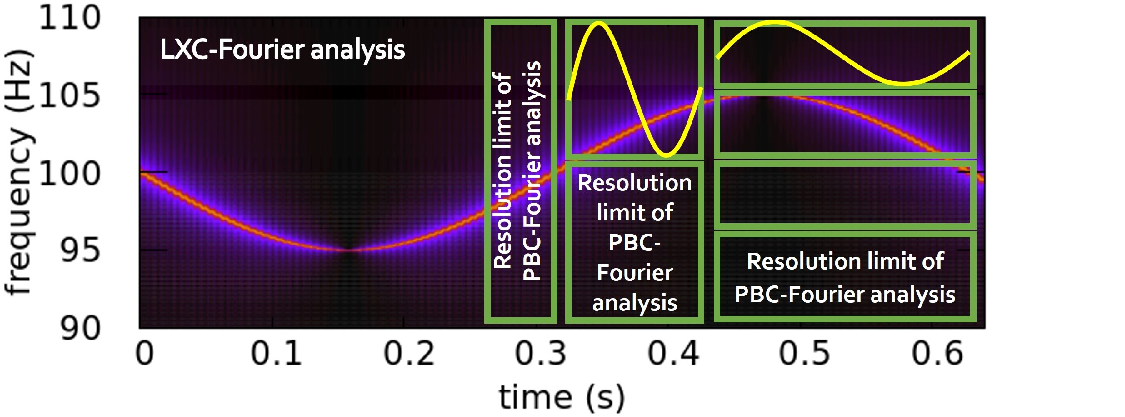}
\caption{
Spectrogram of LXC-Fourier analysis
for FM time series \cite{cspa2025}.
Frequency-width$\times$time-width is always 1~Hz$\cdot$s
for PBC-Fourier analysis.
}
\label{fig-fm}
\end{figure}

As another demonstration, a spectrogram of FM time series using
LXC-Fourier analysis 
is shown in Fig.~\ref{fig-fm}.

Its central frequency is 100~Hz, frequency deviation is $\pm$5~Hz, 
and sampling frequency is 10~kHz \cite{cspa2025}.
Inserted 
pixels
show the time-frequency resolution limit of
conventional 
PBC-Fourier analysis.
Detail of the frequency deviation is not visible with the PBC-Fourier analysis,
because of its resolution limit.

\section{Analysis of lung sounds with abnormalities}

We use a data set of lung sounds \cite{LS-data} with 4~kHz sampling,
which is recorded using a digital stethoscope
with active noise cancellation, 
and with built-in frequency filters $100-500$~Hz.

Various lung sounds are recorded.
Among them, we chose 
the lung sounds with abnormalities
named
``crackles'' with random pulse series,
and 
``wheezing'' with constant frequency.
In addition, we chose a normal lung sound for reference.

\begin{figure}[hbt]
\centering
\includegraphics[width=90mm]{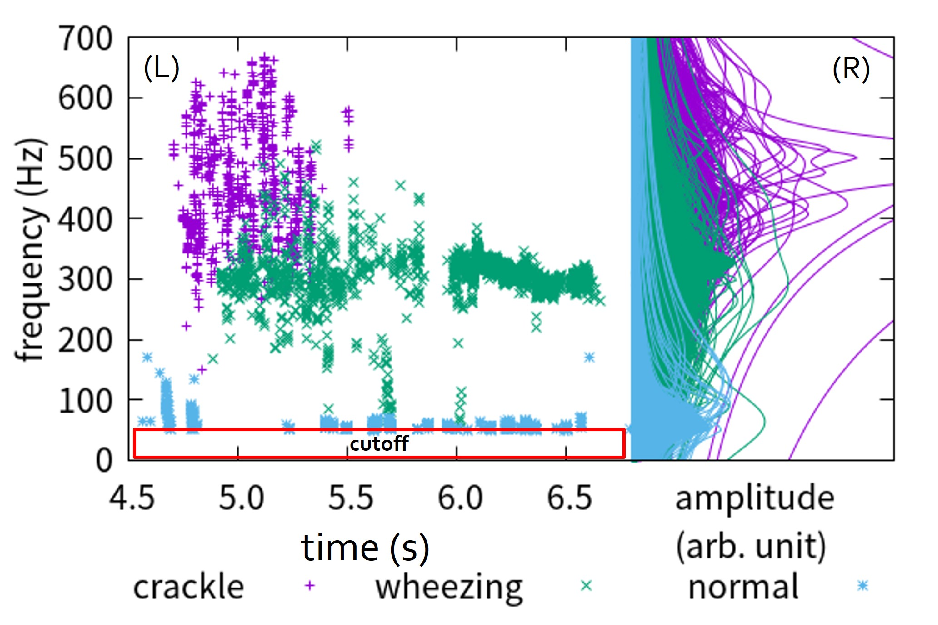}
\caption{
Obtained results for three lung sounds 
(crackles, wheezing, normal).
(L)~Central frequencies, and (R) corresponding instantaneous spectra.}
\label{fig-inst}
\end{figure}

We apply the LXC-Fourier analysis with the parameters of
$M=6$, and $N=20$,
and plot the obtained 
central frequencies $f_m(t)$ above 50~Hz in Fig.~\ref{fig-inst}(L),
and their corresponding instantaneous spectra 
Eq.~(\ref{eq-specg-disc}) in Fig.~\ref{fig-inst}(R).
A frequency filter
$f_m(t)>50$ Hz
is applied to remove the low frequency background terms
below the frequency range of the digital stethoscope,
such an example is 
the constant term (no.~1)
in Fig.~\ref{fig-demo}(b).
This kind of filtering 
(removing unnecessary information)
is valuable to focus on the information which we want.
Note that the computation error term (no. 4) in the figure is also the term unnecessary.

The central frequencies of crackles scatter between $300$ and $700$~Hz, 
and their corresponding instantaneous spectra also scatter around the region.
As the time series of crackles is a pulse series, each pulse decays rapidly, 
each $|\lambda_m(t)|$ becomes large, and each instantaneous spectrum 
becomes broad.

Note that the central frequencies appear above the frequency range of 
the digital stethoscope (up to $500$~Hz).
It is the reflection of high intensities of the pulses,
and some consideration for the frequency range will be required.

The central frequencies of wheezing are around $300$~Hz, 
and their corresponding instantaneous spectra gather around the frequency.

The central frequencies of a normal lung sound 
do not 
appear in the frequency region above $100$~Hz.
This means that the obtained central frequencies for crackles and wheezing
are from abnormalities.

As we applied a frequency filter to remove low frequency background, 
the equation for spectrogram becomes
\begin{equation}\label{eq-specg-disc-flt}
F_{>50}(f,t)=\sum_{m=1}^M 
\sum_{f_m(t)> 50~{\rm Hz}}
\left| \frac{ c_m(t) \lambda_m(t) }{\lambda_m(t)+ 2 \pi i (|f_m(t)|-f)} \right|.
\end{equation}

Various filters are
available for the LXC-Fourier analysis.
Any combinations of filters are available, such as 
$|c_m(t)|>~c_0$ for enough amplitude, 
$|\lambda_m(t)| < \lambda_0$ for narrow spectrum,
$|\lambda_m(t)/f_m(t)| < A_0$ for relatively narrow spectrum,
and
$|c_m(t) f_m(t)| > P_0$ for enough power.
A filter for enough amplitude was also applied to plot Fig.~\ref{fig-inst}(L).

\begin{figure}[thb]
\centering
\includegraphics[width=75mm]{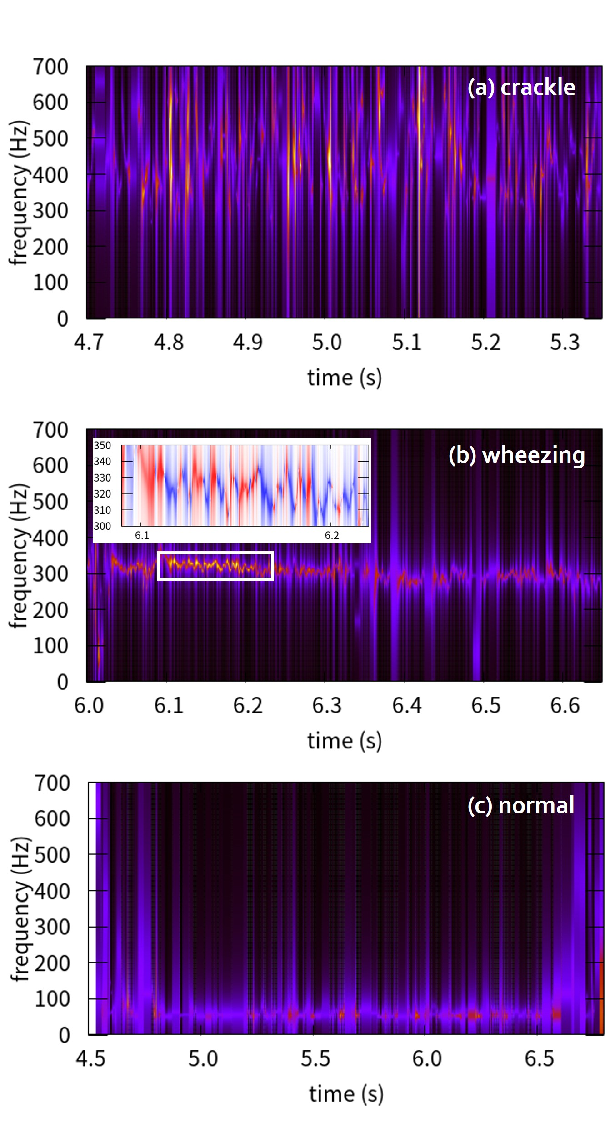}
\caption{
Spectrograms of lung sounds corresponding to Fig.~\ref{fig-inst}.
Interested area are plotted.
(a) Crackles, 
(b) wheezing, 
and
(c) normal.
}
\label{fig-specg}
\end{figure}

The spectrograms corresponding to Fig.~\ref{fig-inst},
using Eq.~(\ref{eq-specg-disc-flt}),
are shown in Fig.~\ref{fig-specg}.
Each interested time region is plotted.

The spectrogram of crackles (Fig.~\ref{fig-inst}(a))
shows intermittent structure, because of its impulsive characteristics.
Each pulse shows a broad spectrum,
and each bright stripe is plotted.

The spectrogram of wheezing (Fig.~\ref{fig-inst}(b)) shows
constant frequency around $300$~Hz.
The insert is the enlargement of the brightest region,
which 
uses Eq.~(\ref{eq-specg-disc-pm}) 
to show additional characteristics.
The growing phases (red) and decaying phases (blue) of the lung sound
are observed
alternately,
which structure is a kind of pulse series.

The spectrogram of a normal lung sound (Fig.~\ref{fig-inst}(c)) shows
no signal above $100$~Hz, and is silent.

Note that the spectrograms with 
conventional 
analysis 
are
found
in Ref.~4.
For example, 
there are no stripes for crackles.

\section{Conclusion}

We proposed to replace
the PBC-Fourier analysis with the LXC-Fourier analysis.
It was shown that 
the proposed LXC-Fourier analysis theoretically includes 
the conventional PBC-Fourier analysis,
and it overcomes the 
``theoretical limit of time-frequency resolution'' of Fourier analysis.

We introduced the method of instantaneous spectra analysis.
Then, we applied the method to 
the lung sounds with abnormalities (crackles and wheezing)
and to a normal lung sound,
as a demonstration.
Among them, crackles contains a random pulse series.

The spectrum of each pulse is available,
and the spectrogram of pulse series is available with assembling each spectrum.
As a result, 
the time-frequency structure of given 
pulse series is visualized.

As the central frequencies of crackles appear above the frequency range of 
the digital stethoscope ($500$~Hz),
which means that the frequency range is insufficient, 
another recording
with enough frequency range
 will be required.

In addition, the time-frequency structure
shown in the insert of Fig.~\ref{fig-specg}(b)
is another 
finding.
This kind of repetitive grow/decay structure is not visible with the conventional
PBC-Fourier analysis,
and
similar structure will be found in various fields
using the LXC-Fourier analysis.

%-------------------------------------------------------------------------

\begin{figure}[thb]
\centering
\includegraphics[width=120mm]{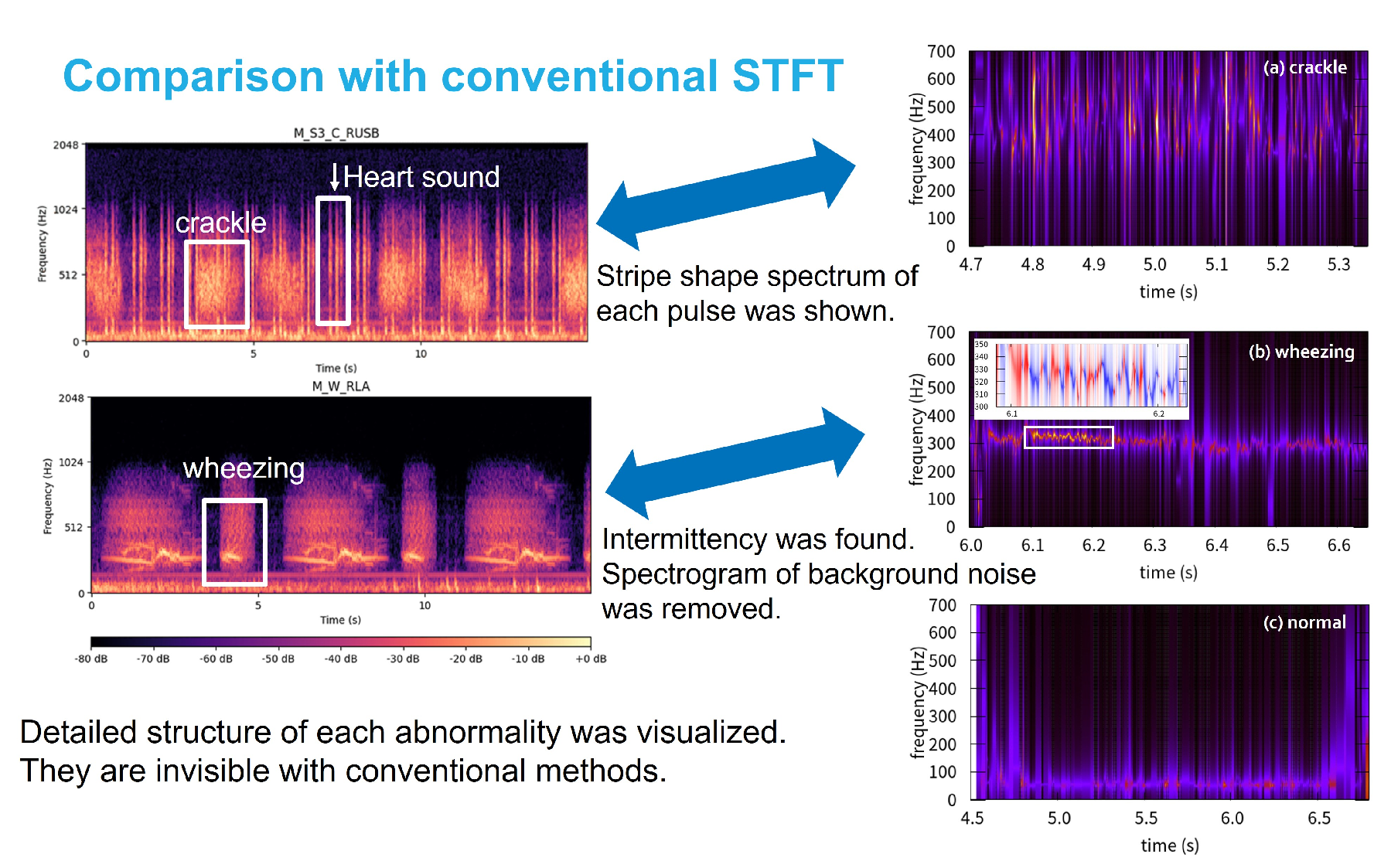}
\caption{
Comparison with conventional method \cite{LS-data}.
}
\label{fig-07}
\end{figure}

\end{document}